\documentclass[betal,sn-mathphys-num]{sn-jnl}

\usepackage{graphicx}%
\usepackage{multirow}%
\usepackage{amsmath,amssymb,amsfonts}%
\usepackage{amsthm}%
\usepackage{mathrsfs}%
\usepackage[title]{appendix}%
\usepackage{textcomp}%
\usepackage{manyfoot}%
\usepackage{booktabs}%
\usepackage{listings}%
\usepackage{svg}%
\usepackage{subcaption}%
\usepackage{algorithm}
\usepackage[noend]{algpseudocode}

\def\defmycaptiontype{on}

\ifthenelse{\equal{\defmycaptiontype}{on}}%
{}%
{\usepackage[labelformat=empty]{caption}}%

\NewDocumentCommand{\mycaption}{m}{%
	\ifthenelse{\equal{\defmycaptiontype}{on}}%
	{\caption{#1}}%
	{\caption[#1]{}}%
}%

\def\defmymarkup{off}
\NewDocumentCommand{\mymarkup}{m}{%
	\ifthenelse{\equal{\defmymarkup}{on}}%
	{{\color{red}#1}}%
	{#1}%
}%

\def\defmysvgtype{pdf}

\NewDocumentCommand{\includemysvg}{O{} m}{%
	\ifthenelse{\equal{\defmysvgtype}{svg}}%
	{\includesvg[#1]{#2}}%
	{\includegraphics[#1]{svg-inkscape/#2_svg-tex.pdf}}%
}%

\makeatletter%
\def\BState{\State\hskip-\ALG@thistlm}%
\algnewcommand\algorithmicforeach{\textbf{for each}}%
\algdef{S}[FOR]{ForEach}[1]{\algorithmicforeach\ #1\ \algorithmicdo}%
\renewcommand{\Function}[2]{\csname ALG@cmd@\ALG@L @Function\endcsname{#1}{#2}\def\jayden@currentfunction{#1}}%
\newcommand{\funclabel}[1]{\@bsphack\protected@write\@auxout{}{\string\newlabel{#1}{{\jayden@currentfunction}{\thepage}}}\@esphack}%
\makeatother%

\theoremstyle{thmstyleone}%
\theoremstyle{thmstyletwo}%

\theoremstyle{thmstylethree}%

\begin{document}

\title[Article Title]{\mymarkup{QuST: QuPath Extension for Integrative Whole Slide Image and Spatial Transcriptomics Analysis}}


%
%
%
%
%
\newbox{\orcid}\sbox{\orcid}{\includegraphics[scale=0.06]{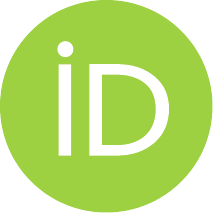}}
\author*[1]{\href{https://orcid.org/0000-0002-3837-8135}{\usebox{\orcid}\hspace{1mm}\fnm{Chao-Hui} \sur{Huang}}}
\author[1]{\href{https://orcid.org/0009-0009-0603-9997}{\usebox{\orcid}\hspace{1mm}\fnm{Sara} \sur{Lichtarge}}}
\author[1]{\href{https://orcid.org/0000-0003-1365-5172}{\usebox{\orcid}\hspace{1mm}\fnm{Diane} \sur{Fernandez}}}
\affil*[1]{\orgname{Pfizer Inc}}



\abstract{\mymarkup{The integration of AI in digital pathology, particularly in whole slide image (WSI) and spatial transcriptomics (ST) analysis, holds immense potential for enhancing our understanding of diseases. Despite challenges such as training pattern preparation and resolution disparities, the convergence of these technologies can unlock new insights. We introduce QuST, a tool that bridges the gap between WSI and ST, underscoring the transformative power of this integrated approach in disease biology.}}




\maketitle

\mymarkup{Spatial analysis, a critical component of pathology, has greatly enhanced our understanding of complex biological processes. Traditional pathology, which involves scrutinizing tissue slides with high-power microscopy, is labor-intensive. However, the advent of digital image analysis (DIA) and machine learning (ML) technologies has broadened the scope of artificial intelligence (AI) in this field. Over the past few decades, a slew of deep learning (DL) based whole slide image (WSI) analysis tools such as QuPath \cite{Bankhead:2017}, TIA Toolbox \cite{Pocock:2022}, MONAI \cite{Cardoso:2022}, SlideFlow \cite{Dolezal:2024}, PHARAOH \cite{Faust:2024}, WSInfer \cite{Kaczmarzyk:2024} have been introduced.}

\mymarkup{One of the significant hurdles in DL-based WSI analysis is the creation of training patterns. Hematoxylin \& eosin (H\&E), the standard tissue staining technique, provides structural information but rarely offers direct biological evidence like gene expressions and transcript factors. As a result, the success of DL-based WSI analysis hinges largely on the expertise of those conducting manual annotation tasks on the WSI H\&E images.}

On the other hand, spatial transcriptomics (ST) has seen significant advancements due to digital pathology, enabling the visualization and analysis of histological sections with gene expression features. ST provides valuable spatial context to molecular data, making it vital for studying complex biological processes, such as cell-cell interactions. ST also presents a unique opportunity to address the challenges of DL-based WSI analysis, as sub-cellular ST technologies are already available. However, merging these two powerful modalities has been difficult due to differences in data formats and analytical methods.

Numerous research studies have delved into the application of tools for the analysis of ST in WSI. For example, Wood \textit{et al.} examined the use of QuPath for image analysis, paired with GeoMx ST, to investigate gene expression variability in colorectal cancer and liver metastases \cite{Wood:2023}. Tippani \textit{et al.} also acknowledged the importance of QuPath as a robust image analysis software \cite{Tippani:2023}. Despite certain constraints, QuPath's significant functionalities in image analysis has been widely recognized. Nonetheless, these studies primarily employ QuPath for initial image analysis, while more detailed ST analysis is conducted using other tools.

\mymarkup{While providers of ST technologies have developed various platforms for visually examining and researching biological insights from given samples, such as Loupe Browser \footnote{\url{https://www.10xgenomics.com/support/software/loupe-browser/latest}}, Xenium Explorer \footnote{\url{https://www.10xgenomics.com/support/software/xenium-explorer/latest}}, and AtoMx Spatial Informatics Platform \footnote{\url{https://nanostring.com/products/atomx-spatial-informatics-platform/atomx-sip-overview/}}, these tools have not fully exploited the functionalities of DIA, resulting in ineffective integration of existing DIA tools and platforms into ST research. To address this gap, we present QuST, a QuPath extension that offers a comprehensive platform for integrating and analyzing Whole Slide Imaging (WSI) and ST data. QuST is designed to enable more in-depth spatial-omics analysis, including cell-cell interactions, cell spatial profiling, and visualization. Furthermore, QuST's implementation of Deep Learning (DL)-based cell categorization and region segmentation methods could facilitate image annotation based on biological evidence.}

\begin{figure*}[tb]
	\centering
	\includemysvg[width = 0.9\linewidth]{diagram}
	\mycaption{QuST workflow includes: (a) users begin by importing ST data into QuPath using QuST. This step may require additional spatial alignment data which can be obtained via FIJI, if the user is working on Xenium dataset (see text). (b) once the ST data is successfully loaded, users can perform analysis and visualization via QuPath and QuST. (c) given the biological evidences provided by ST, users can generate the training set for image based cell classification and region segmentation based on H\&E. Finally, the result generated using the DL module can be further analyzed using the functions described in (b).}
	\label{fig:diagram}
\end{figure*}

\section*{Methods}
\label{sec:methods}

QuST is designed to seamlessly integrate WSI and ST analysis with QuPath, enhancing its capabilities with tools specifically tailored for spatial biology. The extension supports the visualization of spatial gene expression data within the context of histopathological images, enabling users to explore the molecular landscape of tissues at an unprecedented resolution. Below, we will introduce some analyzing tools and use cases available in QuST.


\subsection*{Integrative WSI and ST Analysis at Single-Cell Level}
\label{sec:sptxanal}

Currently, other than the conventional ST format, QuST is able to import data formats including 10x~Genomics Visium, Xenium, nanoString CosMX, \textit{etc.} The major challenge is to align the ST data to WSI due to the fact that different image modalities are involved. For example, in 10x Xenium, the cell localization is based on DAPI staining, while in most cases, the WSIs are H\&E staining. As a result, loading ST data becomes nontrivial.

QuST offers various data loading approaches for different ST data formats, taking into account the need for data alignment. Each ST data format requires a specific approach for proper alignment. For instance, when loading a Xenium dataset with DAPI and H\&E images, we first computed the affine matrix using the SIFT plugin \cite{Lowe:2004} in FIJI \cite{ Schindelin:2012}, as described in Figure~\ref{fig:img_reg}. The approach was an improved version that was optimized based on the guideline provided by 10x~Genomics \footnote{\url{https://www.10xgenomics.com/analysis-guides/he-to-xenium-dapi-image-registration-with-fiji}}. The obtained \textit{affine matrix} and \textit{transfer function} played the key roles for loading single-cell transcriptomics data via QuST.

\begin{figure*}[tbp!]
	\centering
	\begin{subfigure}[t]{0.5\textwidth}
		\centering
		\includegraphics[angle=90,width=0.9\textwidth]{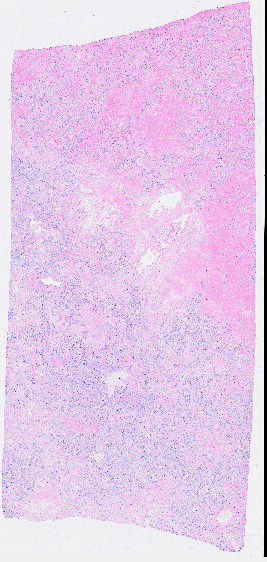}
		\caption{H\&E image}
	\end{subfigure}%
	~
	\begin{subfigure}[t]{0.5\textwidth}
		\centering
		\includegraphics[angle=90,width=0.9\textwidth]{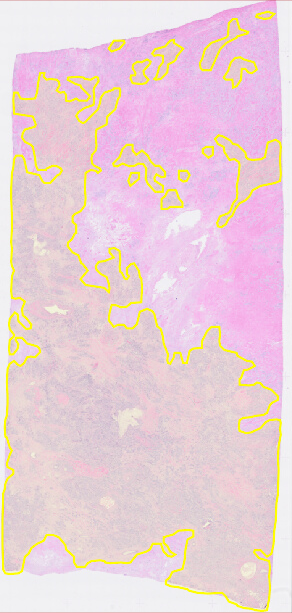}
		\caption{Regions of interest}
	\end{subfigure}%
	\\
	\begin{subfigure}[t]{0.5\textwidth}
		\centering
		\includegraphics[angle=90,width=0.9\textwidth]{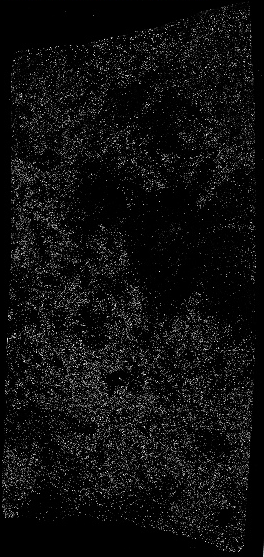}
		\caption{Hematoxylin channel}
	\end{subfigure}%
	~
	\begin{subfigure}[t]{0.5\textwidth}
		\centering
		\includegraphics[angle=90,width=0.9\textwidth]{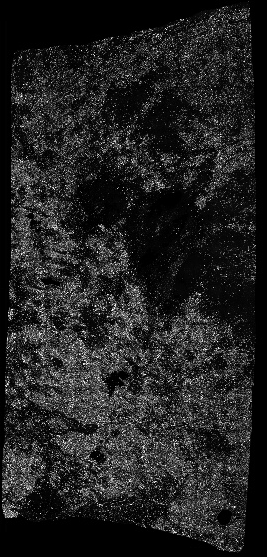}
		\caption{DAPI image}
	\end{subfigure}%
	\\
	\begin{subfigure}[t]{0.5\textwidth}
		\centering
		\includegraphics[angle=90,width=0.9\textwidth]{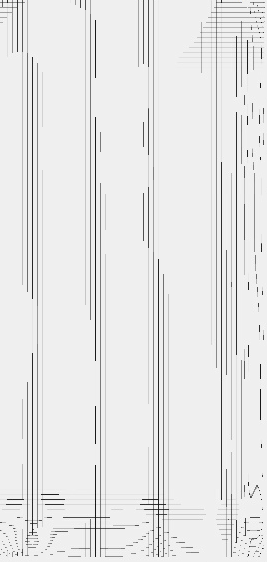}
		\caption{Deformation grid}
		\label{fig:deform_grid}
	\end{subfigure}%
	~%
	\begin{subfigure}[t]{0.5\textwidth}
		\centering
		\includegraphics[angle=90,width=0.9\textwidth]{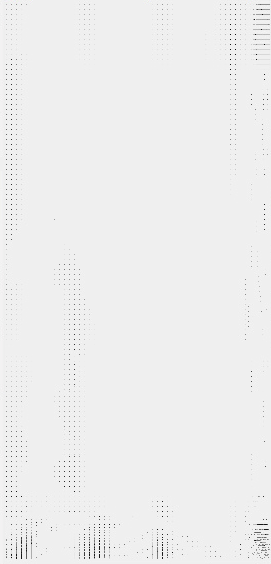}
		\caption{Deformation field}
		\label{fig:deform_field}
	\end{subfigure}%
	\mycaption{\mymarkup{Image example for analyzing the performance of image registration.}}
	\label{fig:img}
\end{figure*}

\begin{figure*}[tbp!]
	\centering
	\begin{subfigure}[t]{0.5\textwidth}
		\centering
		\includegraphics[height=2.5in]{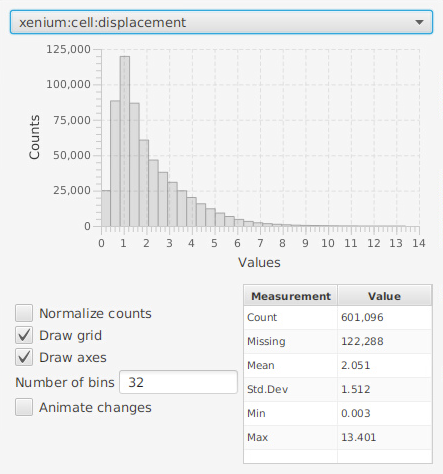}
		\caption{Without image registration}
	\end{subfigure}%
	~%
	\begin{subfigure}[t]{0.5\textwidth}
		\centering
		\includegraphics[height=2.5in]{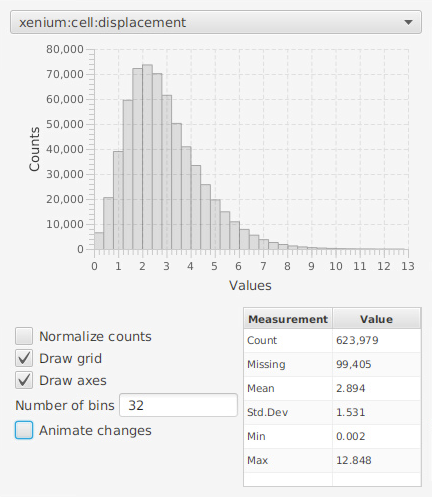}
		\caption{With image registration}
	\end{subfigure}
	\mycaption{\mymarkup{Statistics for cell displacement with and without image registration.}}
	\label{fig:stat}
\end{figure*}

\mymarkup{To test the proposed approach, in the experiment, we first performed a cell detection algorithm, \textit{e.g.}, StarDist \cite{Schmidt:2018}, Cellpose \cite{Stringer:2020}, \textit{etc.} Then, for experimental purpose, we loaded transcriptomic data \textit{with} and \textit{without} including image registration information, separately. The results are shown in Figure~\ref{fig:img}, and the statistical evaluation of cell displacement between the H\&E image and transcriptomic data is shown in Figure~\ref{fig:stat}. In the experiment, 723,384 cells were detected from H\&E images. Without using image registration, 122,288 cells were missing. The root causes include: 1) different quality control approaches for the two data modalities; and 2) the location information obtained from the transcriptomic data did not match the cells detected on the given H\&E images. With image registration, this number dropped to 99,405, representing an 18.71\% improvement. In addition, it can be observed that the deformation looks much relevant to the grid-like artifacts \cite{Wang:2023} resulting from stitching the DAPI images (see Figure~\ref{fig:deform_grid} and \ref{fig:deform_field}). As a result, integrating the proposed linear and nonlinear whole slide image registration can mitigate the noises generated from data acquisition.}

\begin{figure*}[tbp!]
	\centering
	\includemysvg[width = 0.66\linewidth]{imgreg_workflow}
	\mycaption{\mymarkup{Workflow for DAPI-H\&E image registration.}}
	\label{fig:img_reg}
\end{figure*}

\subsection*{Cellular Spatial Profiling}
\label{sec:spatialprofiling}

Cell spatial profiling plays a critical role in spatial-omics analysis. In QuST, cell spatial profiling provides the foundation of all other spatial related computing. First, the Delaunay clustering is required in order to obtain the neighboring cell connectivity. Next, the edge distance of each chosen cells is computed. As a result, the position of each cell in the cluster is obtained and can be used for the following analyzing tasks. The detailed algorithm is shown in Algorithm~\ref{alg:csp}. 

\begin{figure*}[tbp]
	\centering
	\begin{subfigure}[t]{0.5\textwidth}
		\centering
		\includegraphics[width=0.9\textwidth]{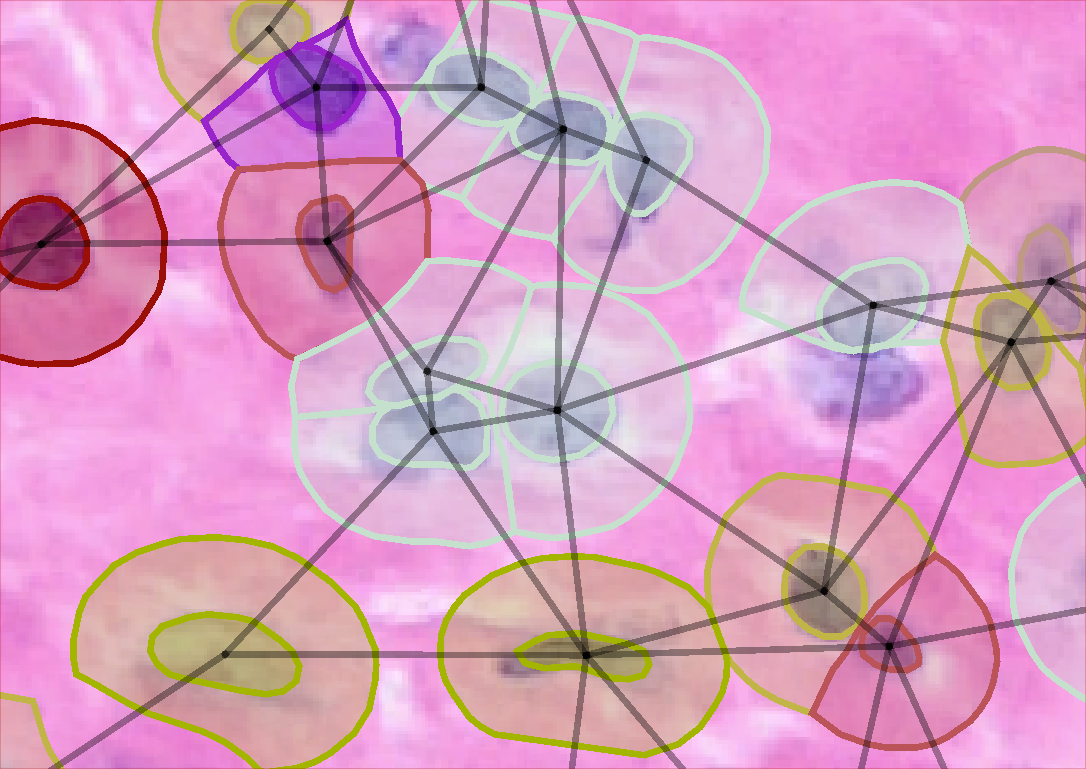}
		\label{fig:result_neighbor_connection}
		\caption{Neighboring cell connectivity}
	\end{subfigure}%
	~%
	\begin{subfigure}[t]{0.5\textwidth}
		\centering
		\includegraphics[width=0.9\textwidth]{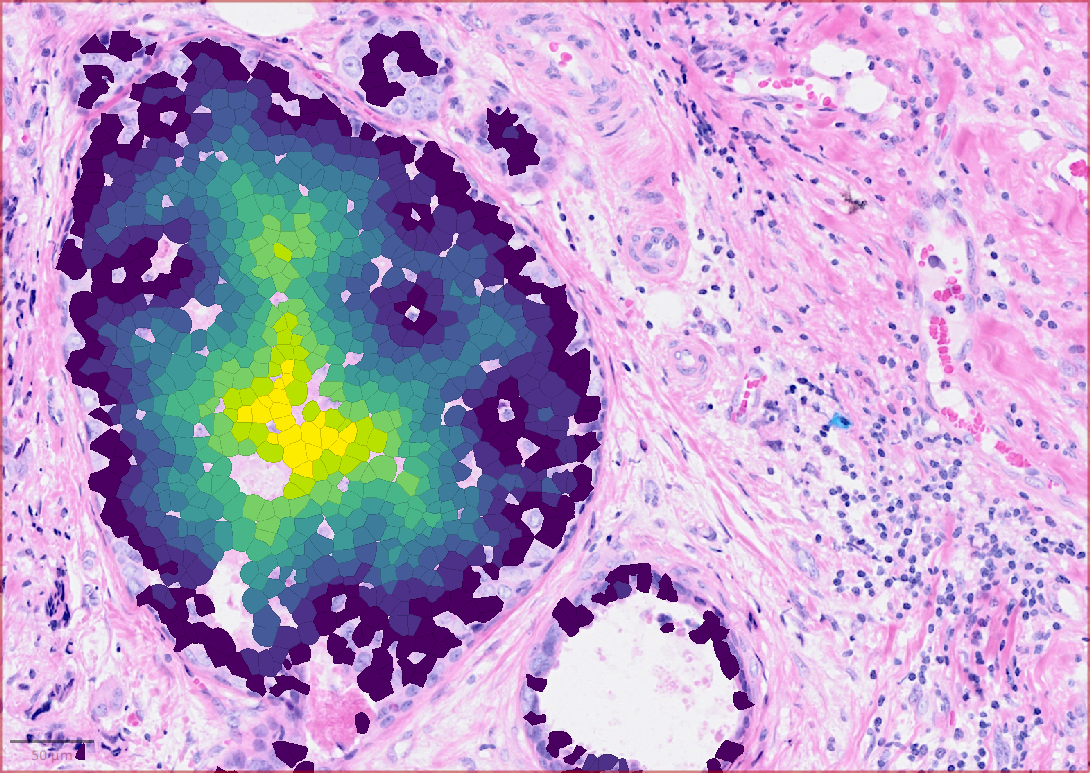}
		\caption{Distance to the boundary of a specific cell-type cluster}
	\end{subfigure}%
	\mycaption{Results showing functions of spatial profiling provided by QuST: (a) Neighboring cell connectivity based on Delaunay clustering. Various single cell analyses available in QuST are based on the neighboring cell connectivity.  (b) QuST's cellular spatial profiling generates a heat map indicating the distance to boundary of a specific cell type, \textit{e.g.}, tumor-epithelial cells to the corresponding tumor boundary.}
	\label{fig:result_spatial}
\end{figure*}

\mymarkup{The results shown in Figure~\ref{fig:result_spatial} represent insights from cellular spatial analysis. The heat map indicates the boundary distance of individual cells, \textit{e.g.,} the distance from a cancer epithelial cell to the boundary of the corresponding tumor boundary. Based on the heat map, one can explore the differential gene expression patterns between the intratumoral tumor cells and the tumor cells present in the immune-invasive region, which are located on the surface of the tumor. }

\mymarkup{A use case of QuST is spatial profiling for tumor micro-environment. The tumor microenvironment encompasses the surrounding cellular and non-cellular components that interact with cancer cells. It plays a crucial role in tumor growth, progression, and response to therapy. By understanding the complex interactions between cancer cells, immune cells, stromal cells, and the extracellular matrix, researchers can identify potential targets for therapeutic intervention. Given the rich information provided by a ST dataset, the functions that QuST can provide are of paramount importance for tumor micro-environment study.}

\begin{algorithm}[tbp]
	\caption{Cell spatial profiling based on Delaunay clustering}\label{alg:csp}
	\begin{algorithmic}[1]
		\Procedure{CellSpatialProfiling}{}
		\State $\mathcal{C} \gets \text{all targeting cells }$
		\ForEach {$c \in \mathcal{C}$}
		\State $\mathcal{N}_{c} \gets \text{all neighbors of }c.$
		\If{$\exists c'\in \mathcal{N}_{c} \text{ where }c'\text{ is not the same category of }c$}
		\State{$e_{c}\gets 0$}\Comment{$e_{c}$, represents edge distance of $c$, is initialized as $0$}
		\EndIf
		\EndFor
		\Statex
		\State {$i\gets 0$}\Comment{$i$: edge distance indicator}
		\Repeat
		\ForEach {$c\in\mathcal{C}$}
		\State {$\mathcal{N}_{c} \gets \text{all neighbors of }c.$}
		\If{$\exists c'\in\mathcal{N}_{c}\text{ where }e_{c'}=i\textbf{ and }\forall c'\in\mathcal{N}_{c},c'\text{ is the same category of }c$}
		\State{$e_{c}\gets i+1$}
		\EndIf
		\EndFor
		\State{$i\gets i+1$}
		\Until{$\forall c\in\mathcal{C},e_{c}$ are obtained}
		\EndProcedure		
	\end{algorithmic}
\end{algorithm}

\subsection*{\mymarkup{Cellular Neighborhood Analysis}}

\mymarkup{Living tissues are composed of various cellular communities that coexist in complex spatial structures. Clusters of cells, each specialized for particular tissues, function together in advanced functional units to uphold and manage organ functions. Thus, the analysis of spatial context is critical for a thorough understanding of tissue biology. As Ruitenberg \textit{et al.} discussed, the development of spatial-omics technologies has equipped us with the means to profile transcriptomes \cite{Ruitenberg:2024}. QuST has offered Cellular Neighborhood Analysis (CNA), which is based on a similar approach to HistoCAT \cite{Schapiro:2017}.}

\begin{figure*}[tbp]%
	\centering%
	\begin{subfigure}[t]{0.5\textwidth}%
		\centering%
		\includegraphics[height=1.75in]{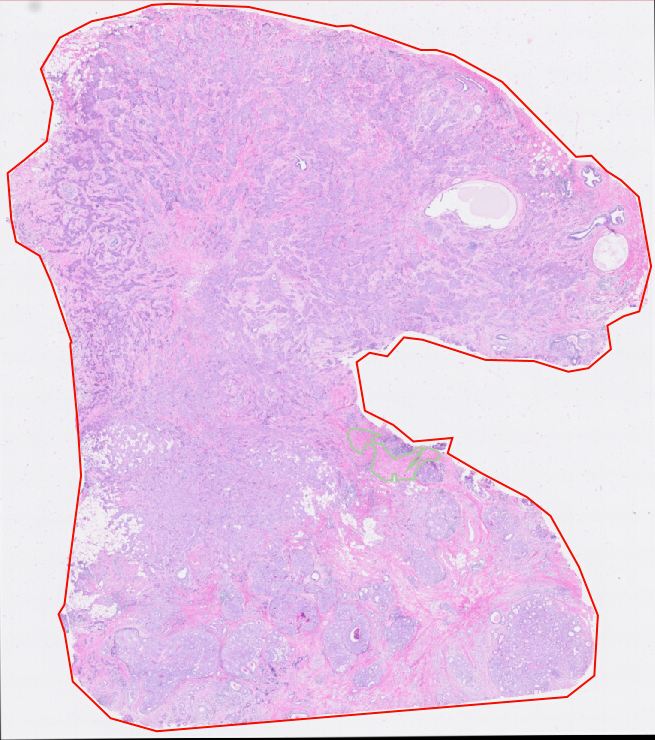}%
		\caption{WSI of the given sample}%
	\end{subfigure}%
	\begin{subfigure}[t]{0.5\textwidth}%
		\centering%
		\includegraphics[height=1.75in]{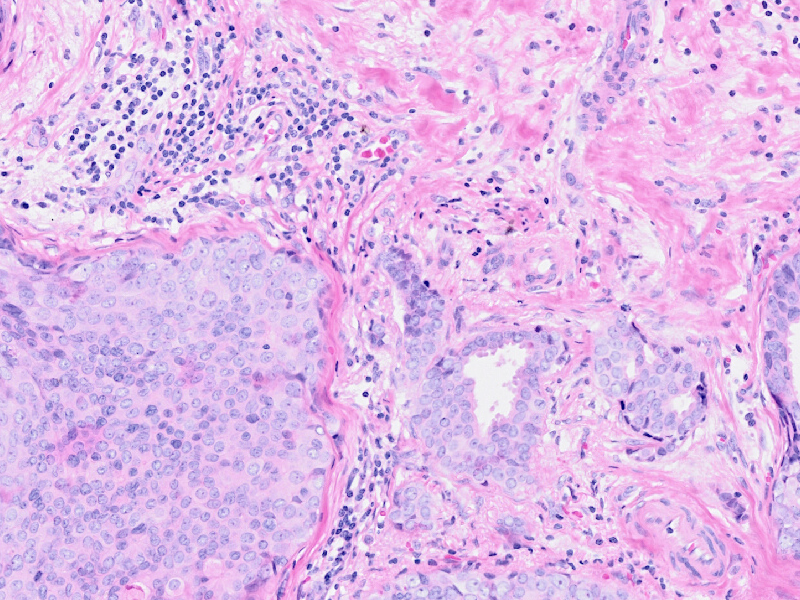}%
		\caption{Magnified ROI}%
	\end{subfigure}%
	\\
	\begin{subfigure}[t]{0.5\textwidth}%
		\centering%
		\includegraphics[height=1.75in]{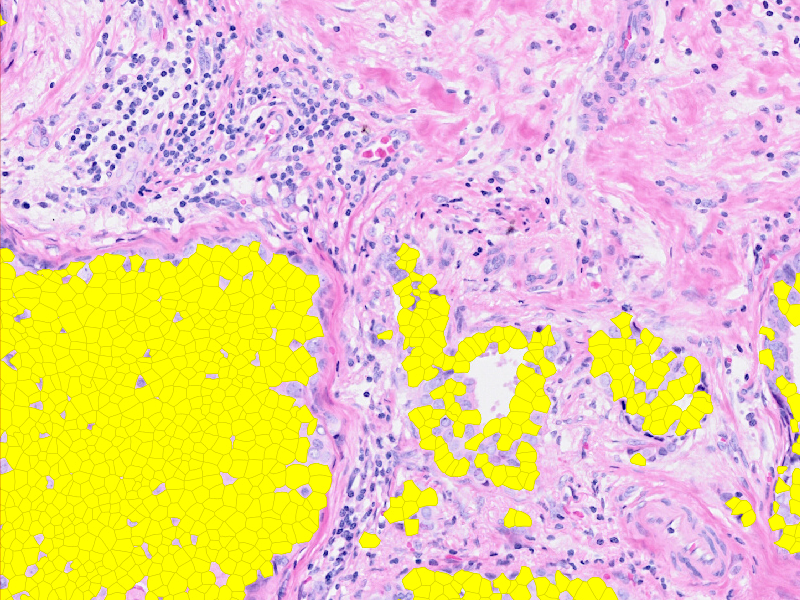}%
		\caption{The labeled regions indicate epithelial and tumor cells}%
	\end{subfigure}%
	~%
	\begin{subfigure}[t]{0.5\textwidth}%
		\centering%
		\includegraphics[height=1.75in]{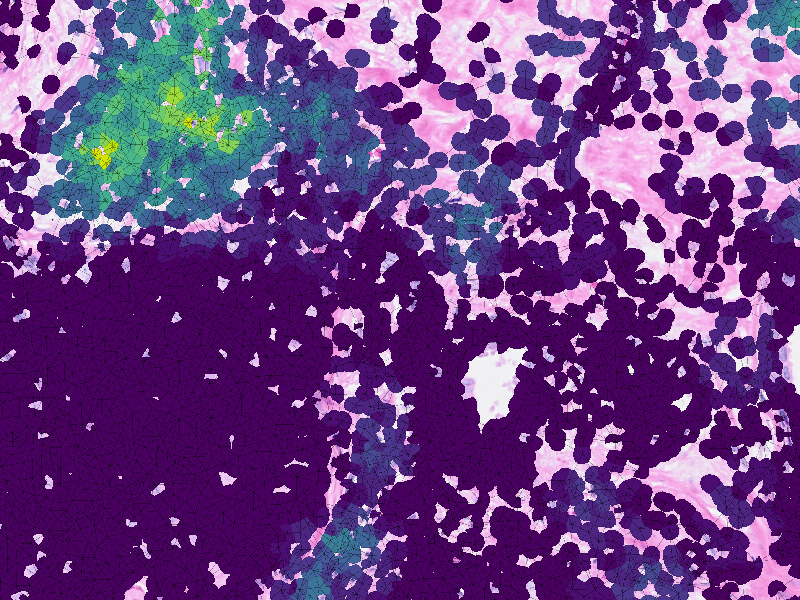}%
		\caption{CNA result.}%
	\end{subfigure}%
	\mycaption{\mymarkup{A result of cellular neighborhood analysis. (a) The given whole slide image. (b) The chosen ROI for the experiment. (c) Epithelial and tumor cells are highlighted as reference. (d) The heap map indicates the number of lymphocytes that can be found in the neighbors of a cell.}}
	\label{fig:result_cna}
\end{figure*}

\mymarkup{An outcome is depicted in Figure~\ref{fig:result_cna}. In the experiment, we initially emphasized the epithelial and tumor cells. We then carried out CNA and used a heat map to display the count of lymphocytes that could be identified in the vicinity of a cell. By overlaying the tumor regions and the CNA results, one can determine the likelihood of a tumor cell having at least one lymphocyte in its immediate surroundings.}

\subsection*{Cell-Cell Interaction Analysis}
\label{sec:cci}

ST is a powerful tool for understanding cell-cell interactions (CCIs) within tissues. By mapping gene expression patterns and spatial organization, researchers gain insights into how cells communicate and influence each other. This knowledge has implications for drug development, disease research, and personalized medicine.

QuST uses the datasets provided by CellTalkDB \cite{Shao:2021}, which is a manual curated database that provides a comprehensive collection of ligand-receptor (LR) pairs in both humans and mice. The database includes 3,398 human LR pairs and 2,033 mouse LR pairs, which were obtained through a combination of text mining, manual verification of known protein-protein interactions using the STRING database, and literature-supported evidence for each pair.

QuST uses the results of \emph{cellular spatial profiling} to compute CCI, effectively incorporating crucial information about cell neighborhoods within specific regions of interest. When analyzing a cell of receptors, QuST takes into account all ligand cells situated within a designated neighboring distance, determined using Delaunay clustering, for the computation of the corresponding CCI. The algorithm presented in Algorithms~\ref{alg:cci} provide detailed explanations of how QuST calculates the LR product. Our future work includes incorporating implementations for more advanced methods \cite{Shao:2021}.

\begin{figure*}[tbp]
	\centering
	\begin{subfigure}[t]{0.5\textwidth}
		\centering
		\includegraphics[height=1.75in]{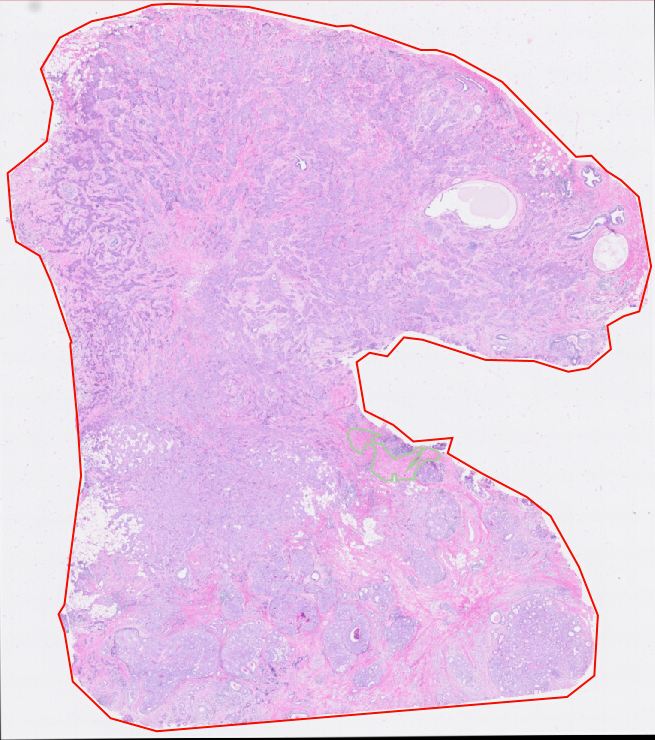}
		\caption{WSI of the given sample}
	\end{subfigure}%
	\begin{subfigure}[t]{0.5\textwidth}
		\centering
		\includegraphics[height=1.75in]{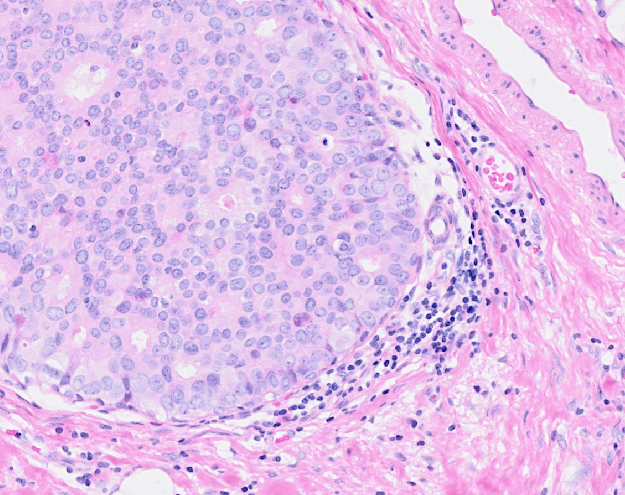}
		\caption{Magnified ROI}
	\end{subfigure}%
	\\
	\begin{subfigure}[t]{0.5\textwidth}
		\centering
		\includegraphics[height=1.75in]{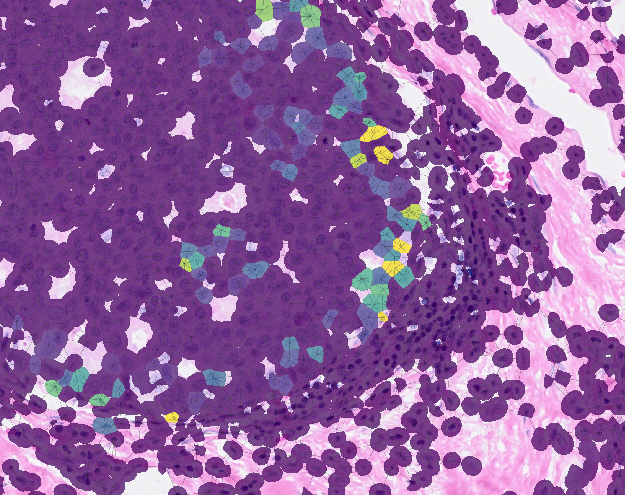}
		\caption{CCI ligand expression}
	\end{subfigure}%
	~%
	\begin{subfigure}[t]{0.5\textwidth}
		\centering
		\includegraphics[height=1.75in]{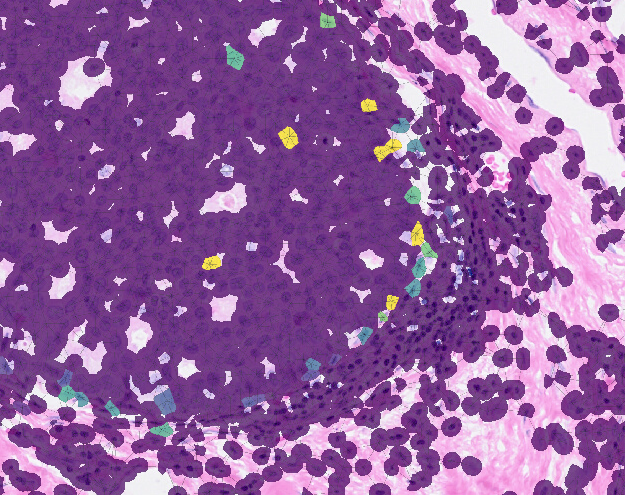}
		\caption{ CCI receptor expression}
	\end{subfigure}%
	\mycaption{\mymarkup{Analyzing CEACAM6-EGFR CCI using QuST: (a) The given WSI. (b) The ROI shows a region of tumor and lymphocyte aggregation. (c) CCI ligand expression. (d) CCI receptor expression.}}
	\label{fig:result_cci}
\end{figure*}

\mymarkup{Figure~\ref{fig:result_cci} shows a result of CCI, focusing on  CEACAM6-EGFR CCI analysis.  As CCI analysis offers an additional layer of investigation by generating a heat map that illustrates the intensity of CCI for a specific ligand-receptor pair. The generated heat map provides a quantitative measure of the strength and significance of communication among a cluster of cells.}

\begin{algorithm}[tbp]
	\caption{An algorithm for computing ligand/receptor expression of CCI using LR product.}
	\label{alg:cci}
	%
	\begin{algorithmic}[1]
		\Procedure{ReceptorCCIProfiling}{$c: \text{targeting cell}$, $l:\text{ligand expression}$, $r:\text{receptor expression}$}
		\State $v_l \gets 0$
		\State $v_r \gets 0$
		\State $\mathcal{N}_{c} \gets \text{all selected neighbors of }c$
		\ForEach {$c' \in \mathcal{N}_{c}$}
		\State $v_r \gets v_r+r\times\Call{GetLigandExpression}{c'}$
		\State $v_l \gets v_l+l\times\Call{GetReceptorExpression}{c'}$
		\EndFor
		\State $v_l\gets v_l/|\mathcal{N}_c|$
		\State $v_r\gets v_r/|\mathcal{N}_c|$\\
		\Return $v_l, v_r$
		\EndProcedure
	\end{algorithmic}
\end{algorithm}

\subsection*{Cell Clustering Analysis}
\label{sec:dbscancellx}

Recent research has highlighted the profound relationship between the functioning of various biological processes, such as cell-cell interactions, and the local densities and positions of cells within cellular monolayers and stratified epithelia. In an attempt to delve deeper into this relationship, K\"{u}chenhoff \textit{et al.} introduced DBSCAN-CellX, a density-based clustering algorithm  \cite{Kuchenhoff:2023}. This algorithm, which is essentially an extension of the Density-Based Spatial Clustering of Applications with Noise (DBSCAN) method initially proposed by Ester \textit{et al.} \cite{Ester:1996}, is specifically tailored to analyze cell localization and tissue physiology. Studying the densities of various cell types and their positioning within cellular monolayers is fundamental to understanding cell interactions and the functioning of diverse biological processes. For instance, Rodriguez \textit{et al.} pointed out that Tertiary Lymphoid Structures (TLS) - irregular congregations of lymphoid cells in inflamed, infected, or cancerous tissues - are associated with a more favorable cancer prognosis \cite{Rodriguez:2021}. Moreover, Barmpoutis \textit{et al.} proposed a technique for identifying TLS and evaluating their density on H\&E-stained digital slides of lung cancer \cite{Barmpoutis:2021}. As a result of these advancements, DBSCAN-CellX has been successfully integrated into QuST.

\mymarkup{To facilitate cell clustering analysis, we integrated DBSCAN-CellX into QuST. Given the classes of the cells, QuST is able to compute local densities and positions of cells accordingly. The results are shown as the measurements in the detection table. Hence, the results can also be visually investigated and exported using the native QuPath functions of \emph{measurement maps}.}

\mymarkup{The result is shown in Figure~\ref{fig:dbscan_cellx_lympho_example}, indicating that clusters of lymphocytes (50+) were identified from the given sample. This integration allows researchers in oncology to utilize DBSCAN-CellX directly for quantitatively analyzing the morphology of lymphocyte clustering, potentially identifying imaging biomarkers that correlate with disease prognosis.}

\begin{figure*}[tbp!]
	\centering
	\includemysvg[width=0.9\textwidth]{dbscan_cellx_lympho_example}
	\mycaption{\mymarkup{An example of performing DBSCAN-CellX for lymphocytes. (left) Clusters of lymphocytes (50+) were identified from this example. (right) The color annotated on a selected cell indicates the ``edge degree'' of the cell, which refers to the distance to the edge of the corresponding lymphocyte cluster.}}
	\label{fig:dbscan_cellx_lympho_example}
\end{figure*}

\subsection*{Pseudo Spot Generation based on Single Cell ST Data}

\label{sec:pseudospotgeneration}

While sub-cellular ST technologies exist, they often have limitations in terms of the range of the genome they can cover. Consequently, lower-resolution technologies capable of analyzing the entire genome continue to be widely used. As a result, generating data for evaluating gene expression deconvolution approaches is required. 

\begin{figure*}[tbp]
	\centering
	\begin{subfigure}[t]{0.5\textwidth}
		\centering
		\includegraphics[width=0.9\textwidth]{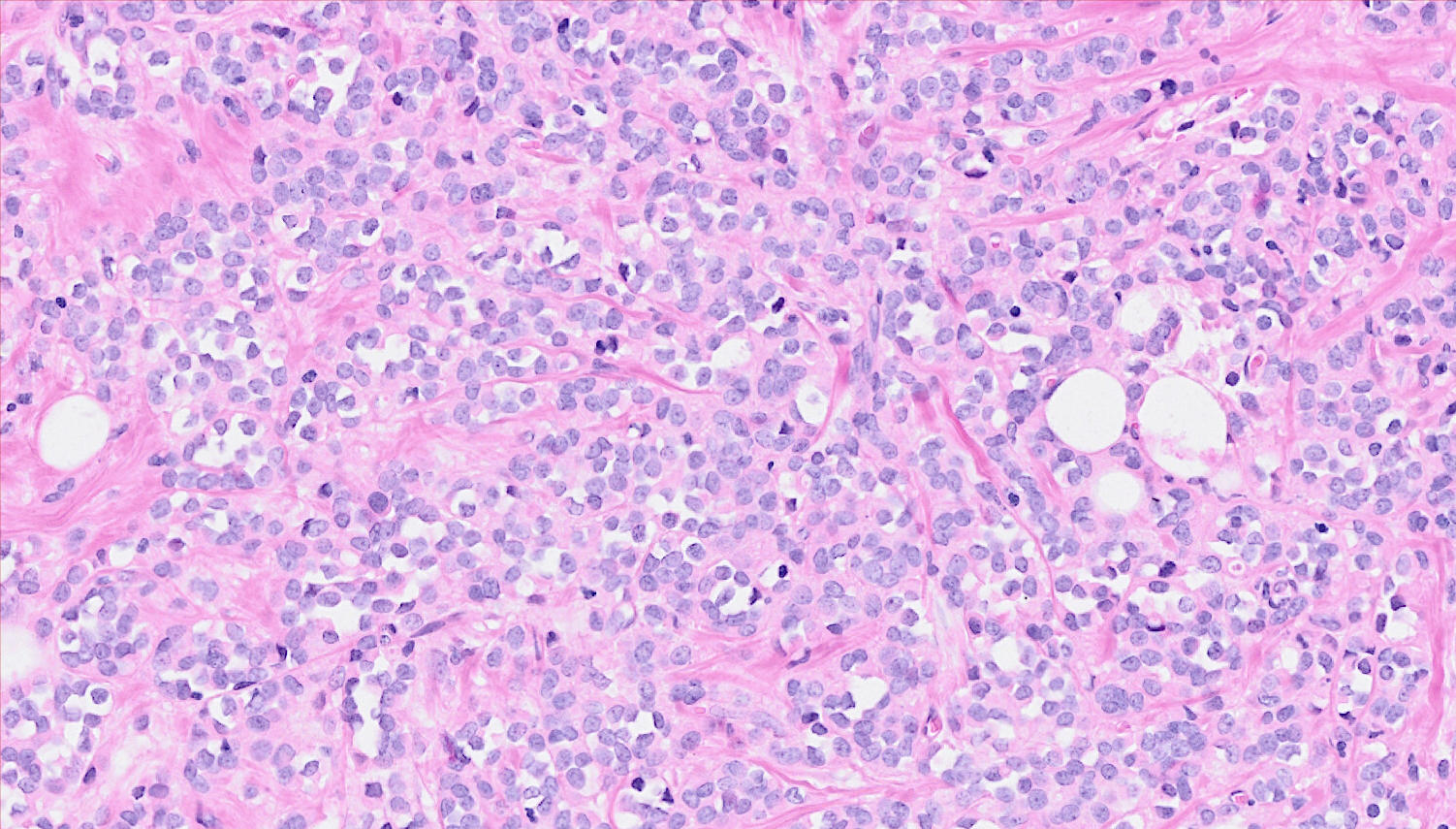}
		\caption{The chosen ROI.}
	\end{subfigure}%
	\begin{subfigure}[t]{0.5\textwidth}
		\centering
		\includegraphics[width=0.9\textwidth]{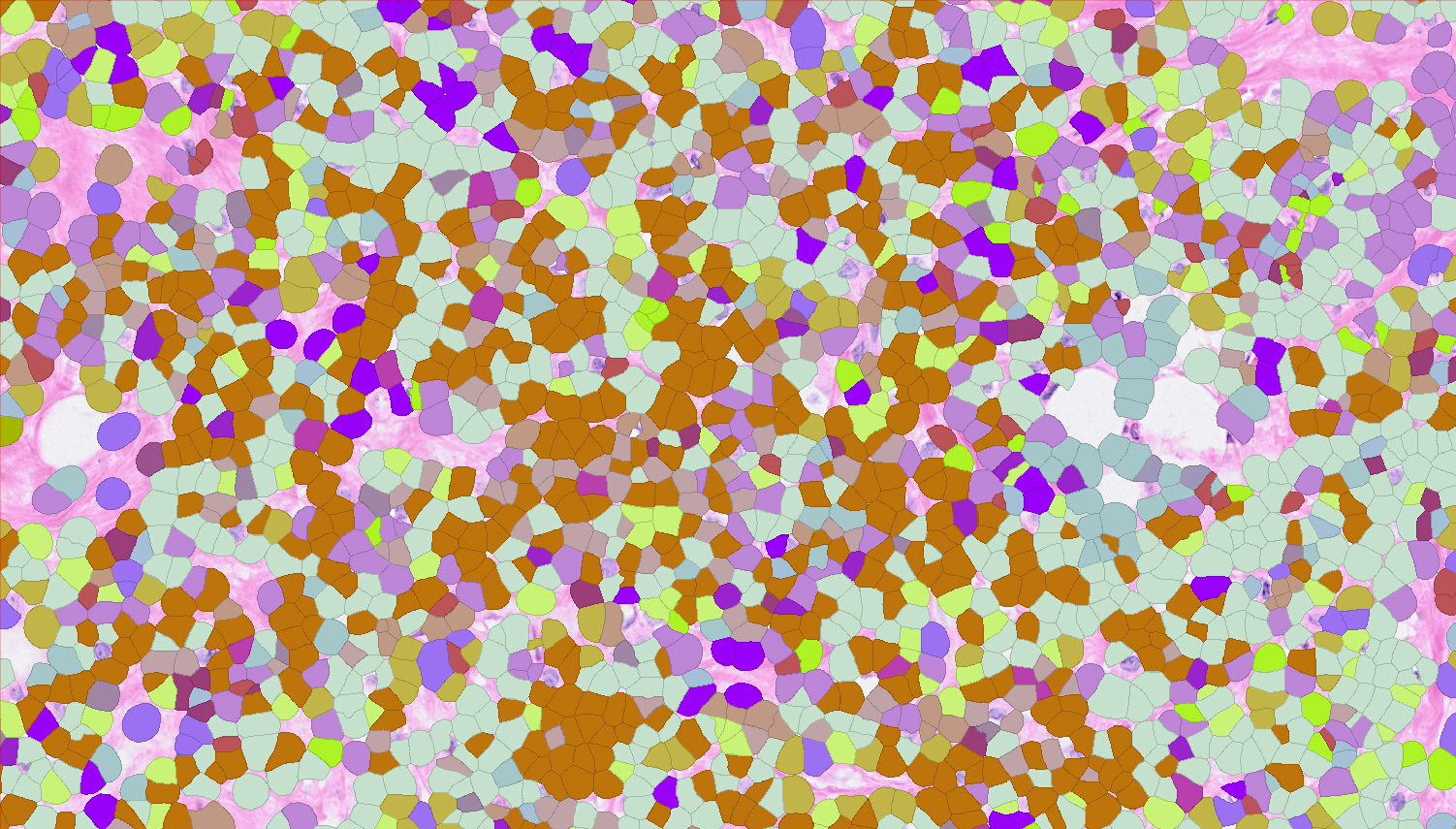}
		\caption{Single cell data.}
	\end{subfigure}%
	\\
	\begin{subfigure}[t]{0.5\textwidth}
		\centering
		\includegraphics[width=0.9\textwidth]{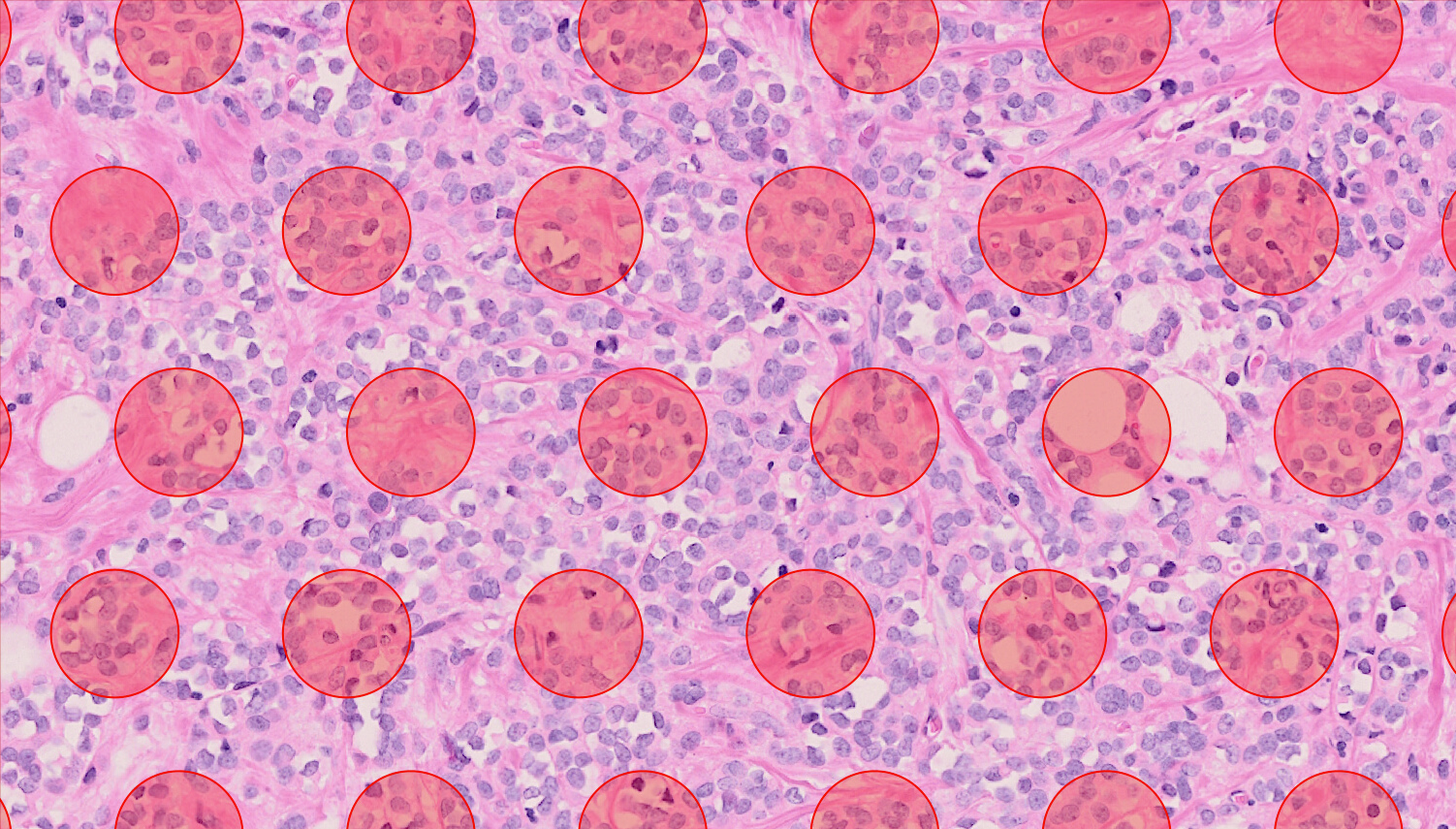}
		\caption{Generated pseudo spots.}
	\end{subfigure}%
	~%
	\begin{subfigure}[t]{0.5\textwidth}
		\centering
		\includegraphics[width=0.9\textwidth]{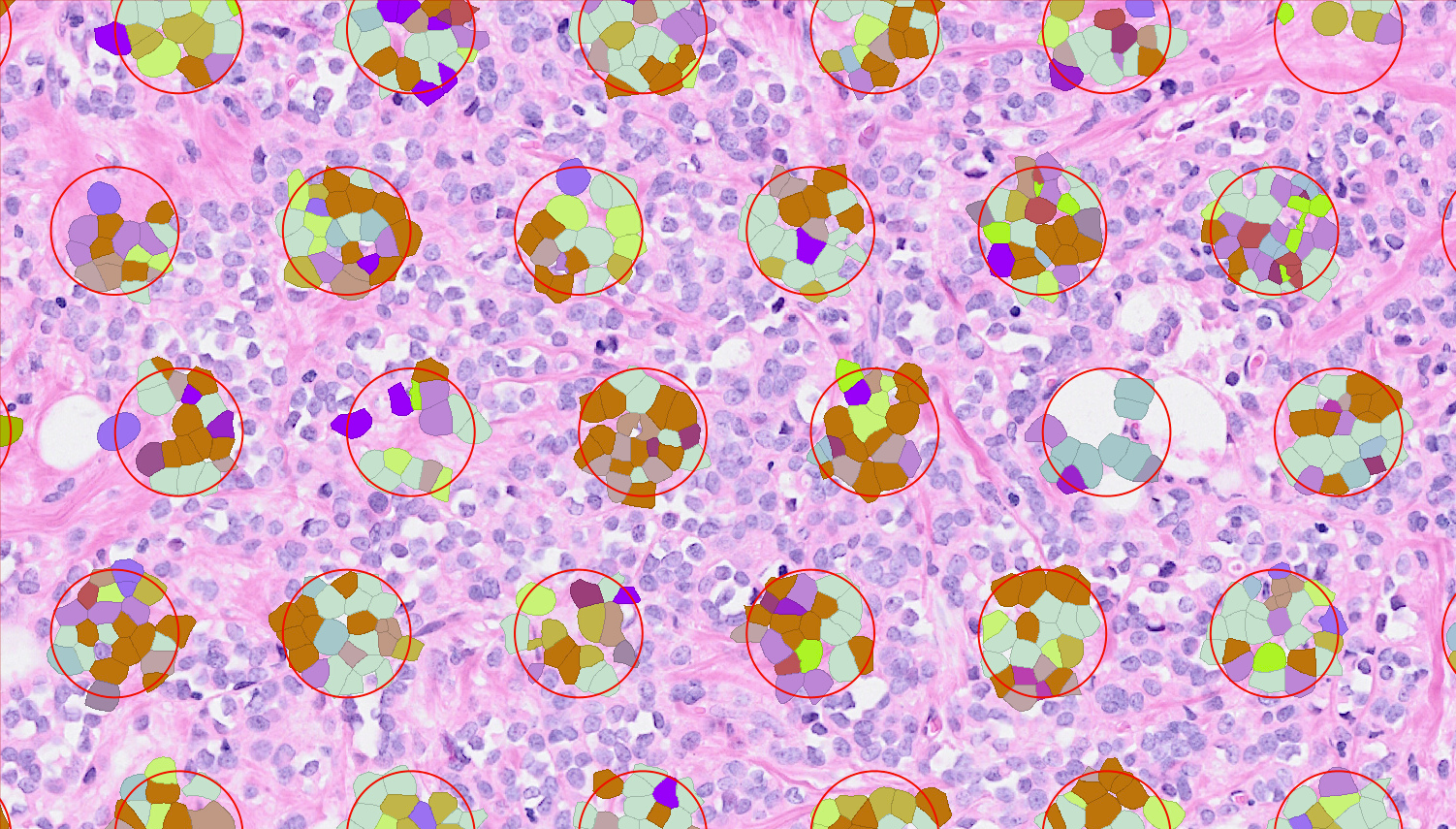}
		\caption{Pseudo spots containing single cell data.}
	\end{subfigure}%
	\mycaption{\mymarkup{Pseudo spots generated using single-cell data.}}
	\label{fig:result_pseudo}
\end{figure*}

QuST  provides an opportunity to evaluate the spatial single cell deconvolution methods by mimicking the Visium datasets (see Figure~\ref{fig:result_pseudo}). For example, Huang \textit{et al.} proposed an approach of spatial transcriptome auto-encoder and deconvolution method, largely utilizing the integration of a scalable deep generative model for predicting gene expression at cellular or nuclei level based on H\&E imaging and \textit{in situ} RNA capturing , thus allowing a better understanding of the tissue micro-environment \cite{Huang:2023b}. 

\subsection*{Image-based Object Classification and Region Segmentation}

QuST embodies a distinctive ability, serving as a link between the realms of transcriptomics and imaging. Its primary utility lies in utilizing transcriptomic data as the biological basis for training deep learning-based object categorization and region segmentation for H\&E images. To service this purpose, QuST features functions for extracting single-cell H\&E images for cell classification, as well as whole slide image patches for region segmentation. The QuPath annotation and detection measurement tables can be exported as files, serving as image label data for DL model training. Consequently, QuST significantly reduces the workload for users training their models for cell classification and region segmentation.

QuST includes a DL capability based on PyTorch \cite{AutomaticdifferentiationinPyTorch} to perform image classification, which aids in cell categorization and region segmentation tasks. To initiate the training procedure, two inputs are required. Firstly, an image set in a folder generated using the aforementioned image sampling approach. Secondly, the detection/annotation measurement table, which can be directly generated from QuPath measurement functions. QuST offers a wide range of neural networks, including resnet \cite{He:2015}, vgg \cite{Simonyan:2014}, densenet \cite{Huang:2016}, and various variations of modern vision transformers (ViT) \cite{Dosovitskiy:2020}. 

\paragraph{\mymarkup{Object Classification}}

\begin{figure*}[h!]
	\centering
	\includegraphics[width=0.9\textwidth]{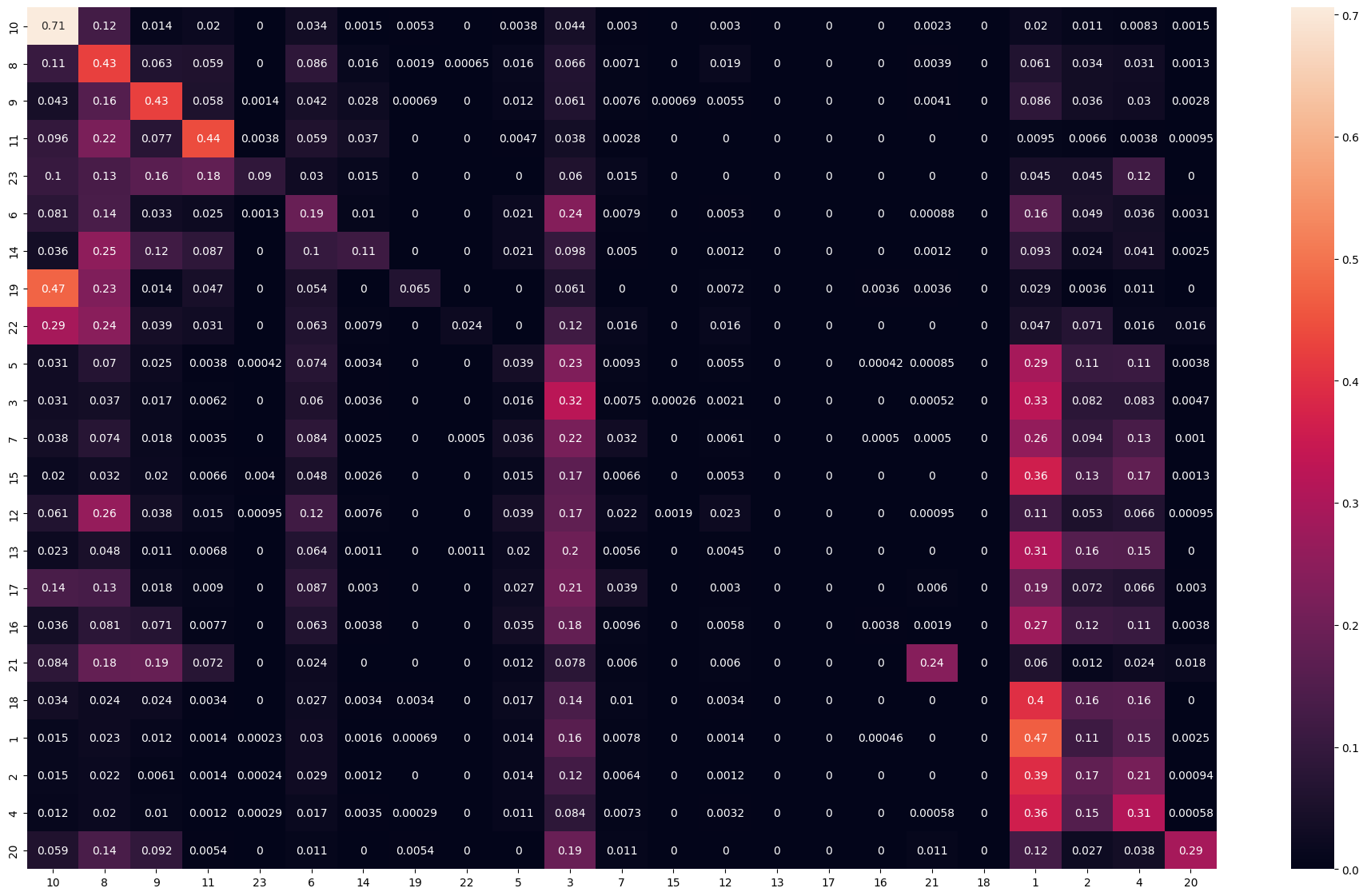}
	\mycaption{\mymarkup{Confusion matrix of single-cell genotype classification based on H\&E.}}
	\label{fig:result_objcls_cm}
\end{figure*}

\begin{figure*}[tbp]
	\centering
	\begin{subfigure}[t]{0.33\textwidth}
		\centering
		\includegraphics[width=0.9\textwidth]{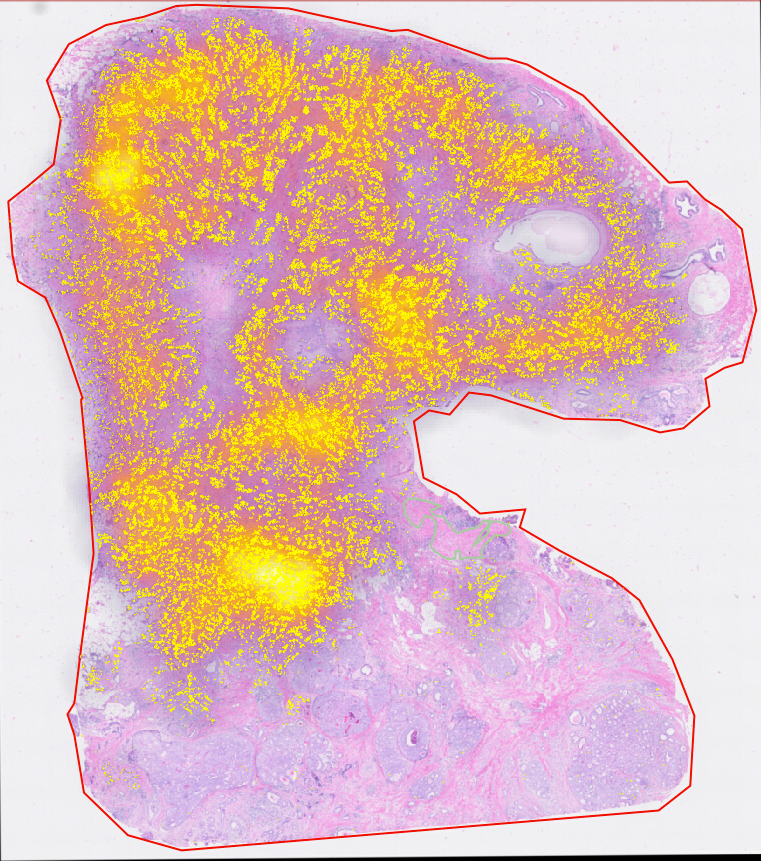}
		\caption{Ground truth (type: 1)}
	\end{subfigure}%
	~%
	\begin{subfigure}[t]{0.33\textwidth}
		\centering
		\includegraphics[width=0.9\textwidth]{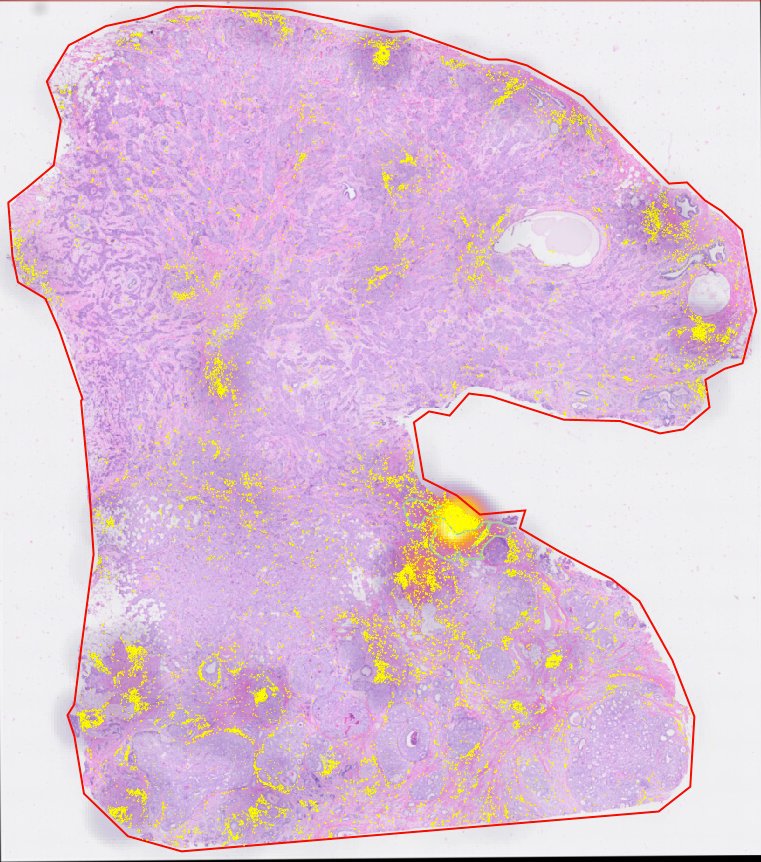}
		\caption{Ground truth (type: 10)}
	\end{subfigure}%
	~%
	\begin{subfigure}[t]{0.33\textwidth}
		\centering
		\includegraphics[width=0.9\textwidth]{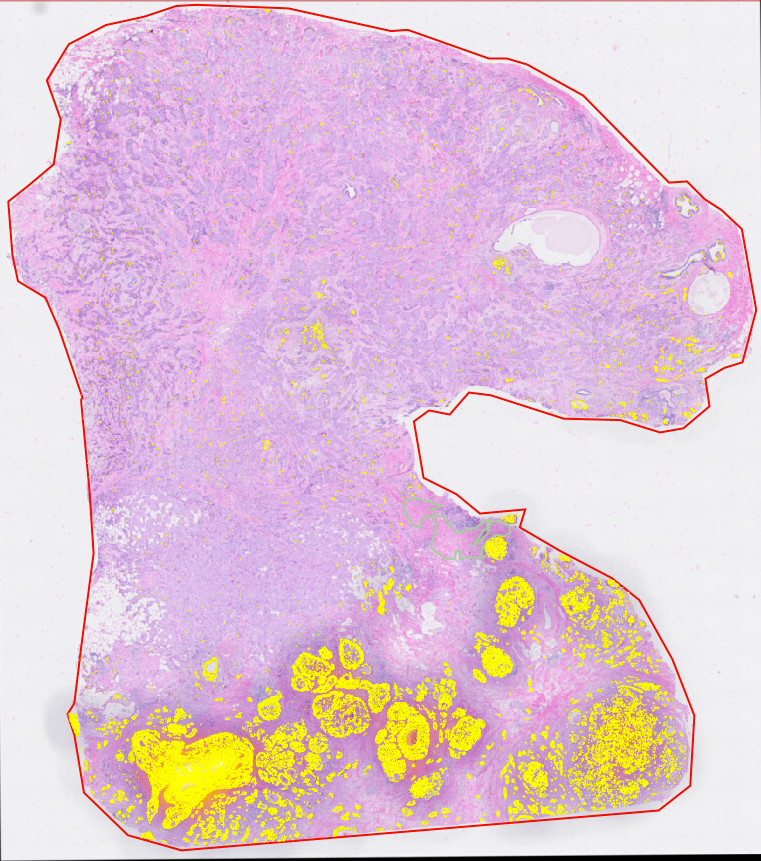}
		\caption{Ground truth (type: 4)}
	\end{subfigure}%
	\\
	\begin{subfigure}[t]{0.33\textwidth}
		\centering
		\includegraphics[width=0.9\textwidth]{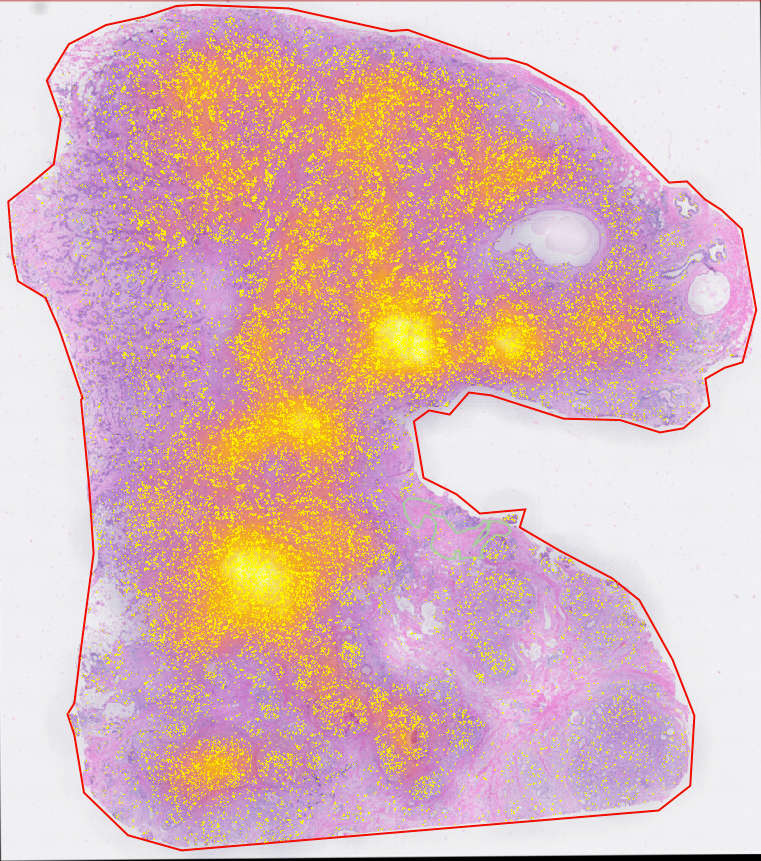}
		\caption{Prediction (type: 1)}
	\end{subfigure}%
	~%
	\begin{subfigure}[t]{0.33\textwidth}
		\centering
		\includegraphics[width=0.9\textwidth]{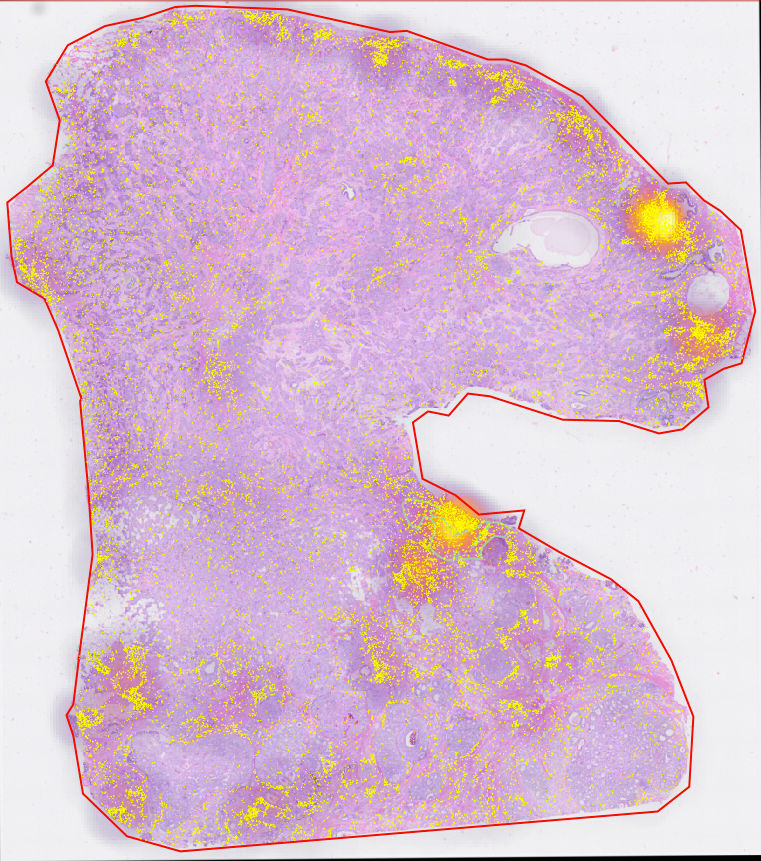}
		\caption{Prediction (type: 10)}
	\end{subfigure}%
	~%
	\begin{subfigure}[t]{0.33\textwidth}
		\centering
		\includegraphics[width=0.9\textwidth]{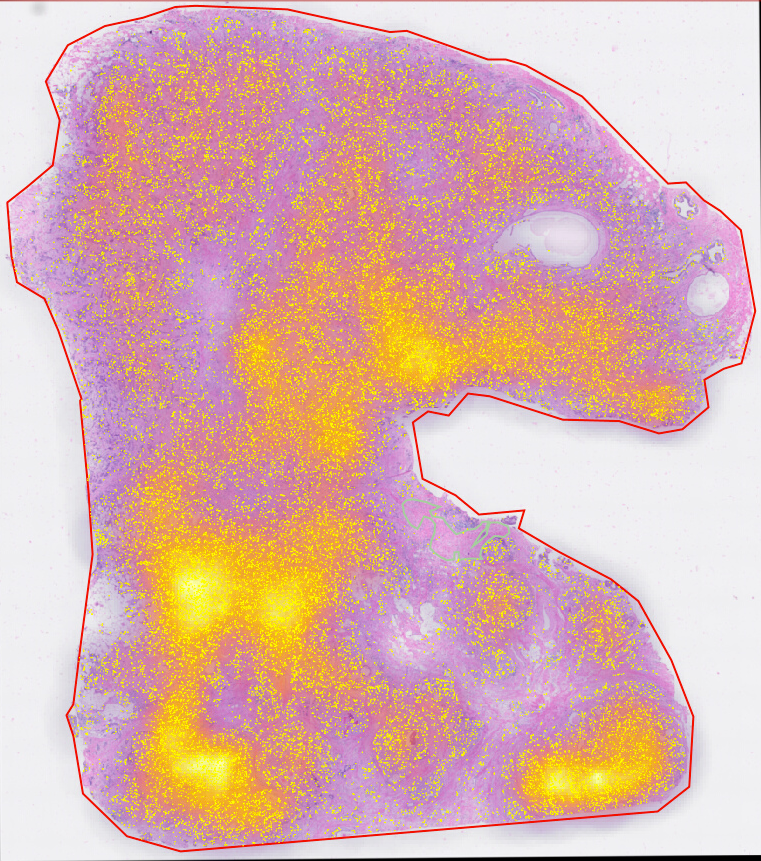}
		\caption{Prediction (type: 4)}
	\end{subfigure}%
	\caption{\mymarkup{Some examples of cell genotype classification: (a-c) The ground truth samples. (d-f) The predicted outcomes.}}
	\label{fig:result_objcls}
\end{figure*}

\mymarkup{In the experiment, we first acquired H\&E image patches for each detected object (\textit{e.g.}, a cell). Next, we used the genotype information provided by the chosen datasets, and train the DL model for object classification. In this experiment, we used ViT. A confusion matrix shown in Figure~\ref{fig:result_objcls_cm}, which was computed based on 10 fold cross-validation. Some examples of single-cell genotype classification based on H\&E are shown in Figure~\ref{fig:result_objcls}. In our experiment, as shown in the confusion matrix, cell type 1 and 10 can be better predicted based on single-cell H\&E images, while type 4 has a poor prediction result. This result revealed that certain cell types are more readily distinguishable in H\&E staining, such as lymphocytes, blood cells, \textit{etc.}, while others are not. Furthermore, differentiating B-cells from T-cells based on H\&E staining is recognized as a particularly challenging task.}

\paragraph{\mymarkup{Region Segmentation}}

\begin{figure*}[h!]
	\centering
	\begin{subfigure}[t]{0.5\textwidth}
		\centering
		\includegraphics[angle=0,width=0.9\textwidth]{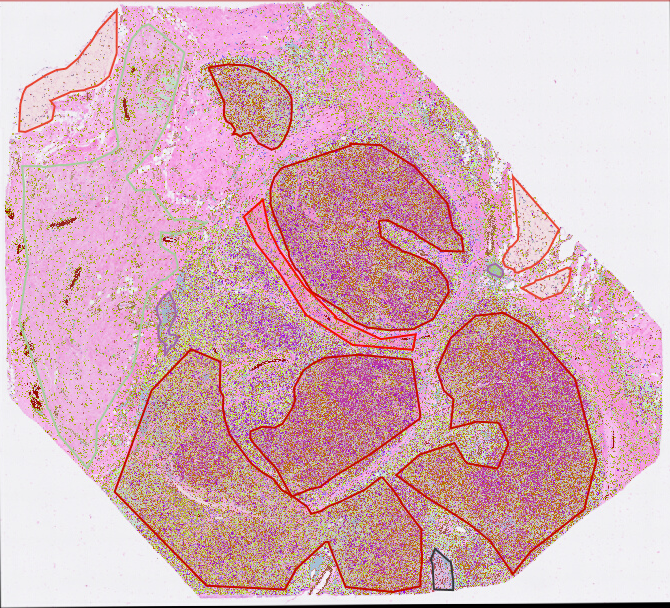}
		\caption{H\&E image as the training set}
		\label{fig:result_regseg_trainingset}
	\end{subfigure}%
	\\
	\begin{subfigure}[t]{0.33\textwidth}
		\centering
		\includegraphics[angle=0,width=0.9\textwidth]{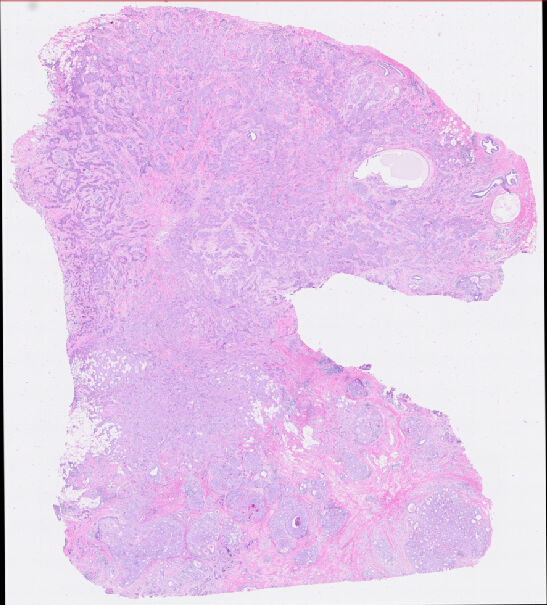}
		\caption{H\&E image testing image}
		\label{fig:result_regseg_testset}
	\end{subfigure}%
	~%
	\begin{subfigure}[t]{0.33\textwidth}
		\centering
		\includegraphics[angle=0,width=0.9\textwidth]{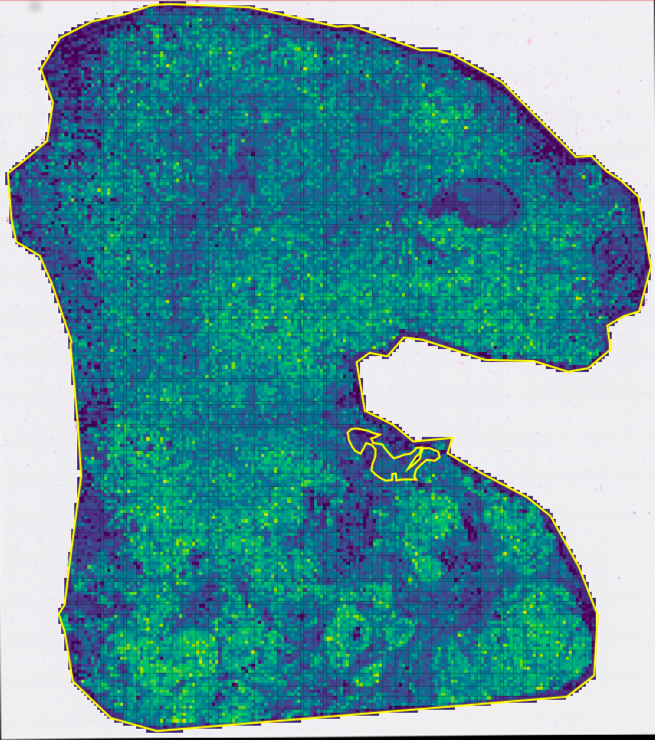}
		\caption{Tumor regions}
		\label{fig:result_regseg_tumor_prob}
	\end{subfigure}%
	~%
	\begin{subfigure}[t]{0.33\textwidth}
		\centering
		\includegraphics[angle=0,width=0.9\textwidth]{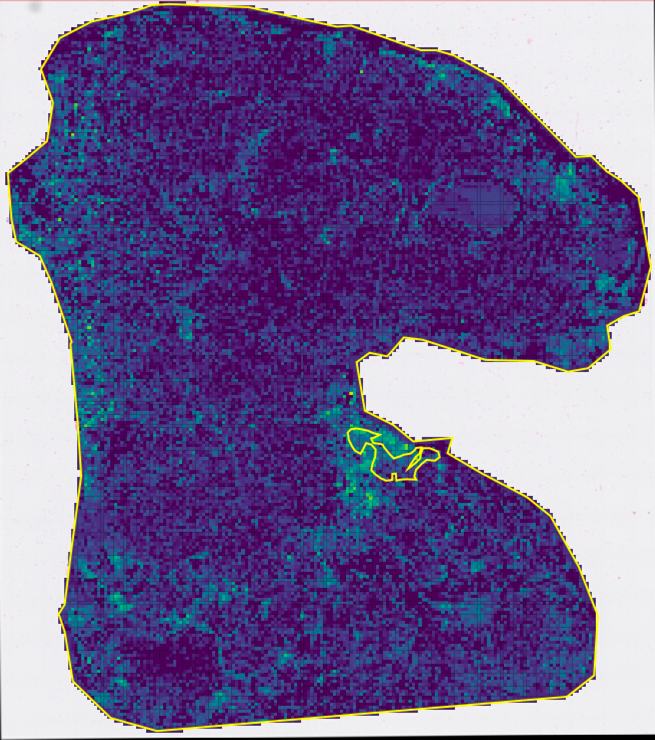}
		\caption{Immuno regions}
		\label{fig:result_regseg_immuno_prob}
	\end{subfigure}%
	\\%
	\begin{subfigure}[t]{0.33\textwidth}
		\centering
		\includegraphics[angle=0,width=0.9\textwidth]{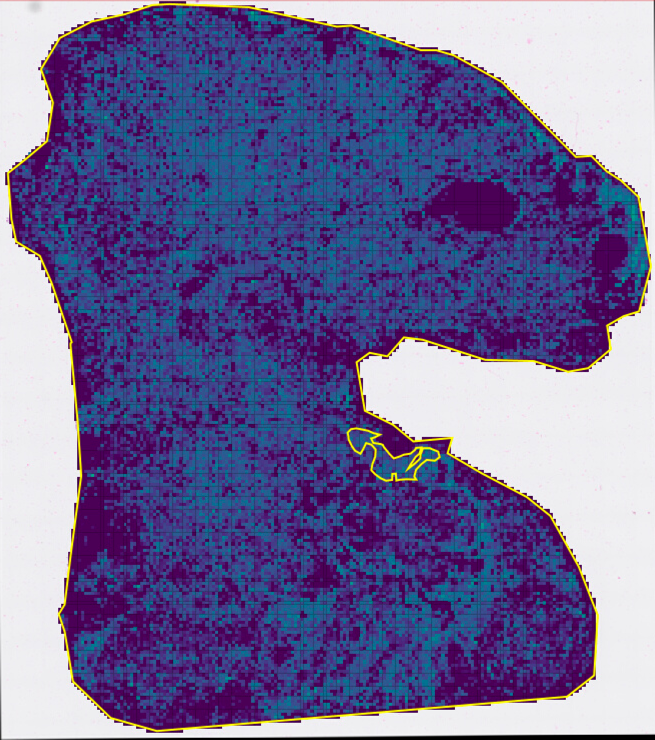}
		\caption{Tumor regions}
		\label{fig:result_regseg_stroma_prob}
	\end{subfigure}%
	~%
	\begin{subfigure}[t]{0.33\textwidth}
		\centering
		\includegraphics[angle=0,width=0.9\textwidth]{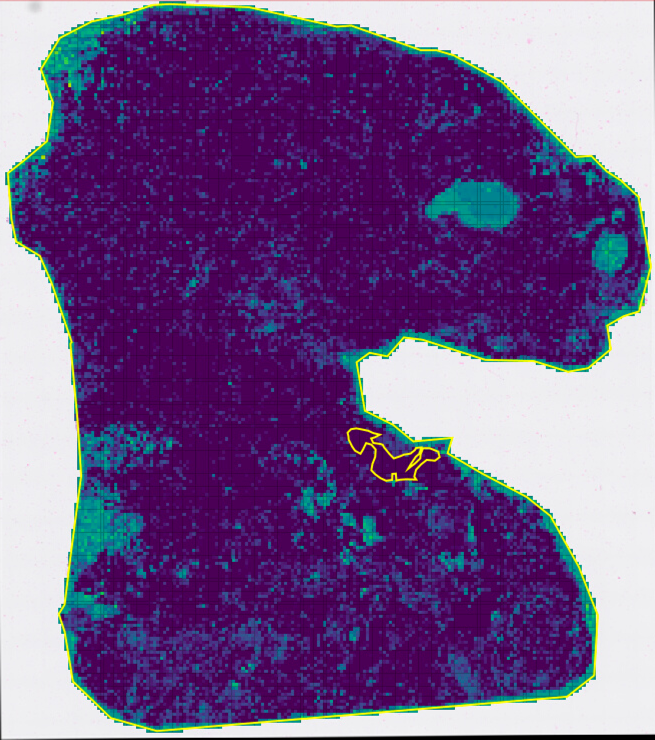}
		\caption{Adipose regions}
		\label{fig:result_regseg_adipose_prob}
	\end{subfigure}%
	~%
	\begin{subfigure}[t]{0.33\textwidth}
		\centering
		\includegraphics[angle=0,width=0.9\textwidth]{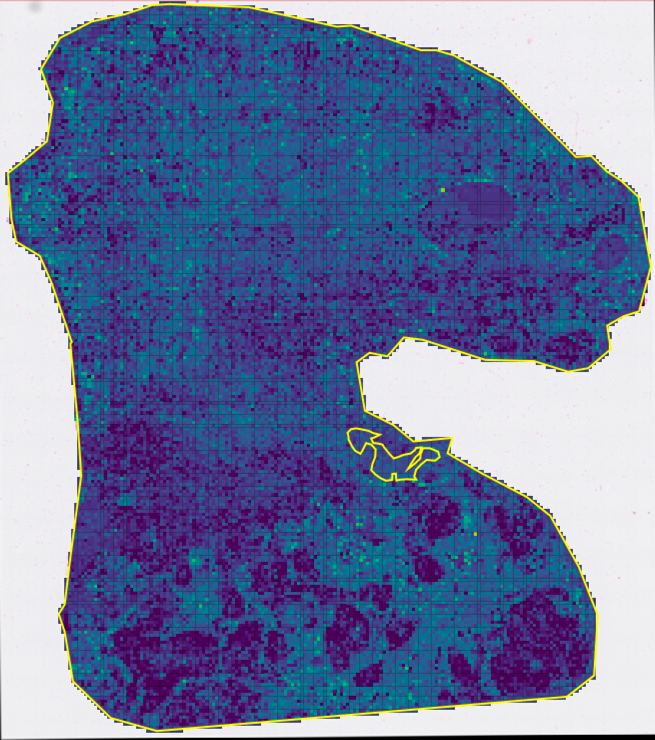}
		\caption{Necrosis regions}
		\label{fig:result_regseg_necrosis_prob}
	\end{subfigure}%
	\mycaption{\mymarkup{Results showing the heap map representing the probability of various classification result.}}
	\label{fig:result_regseg}
\end{figure*}

\begin{figure*}[h!]
	\centering
	\includegraphics[height=3in]{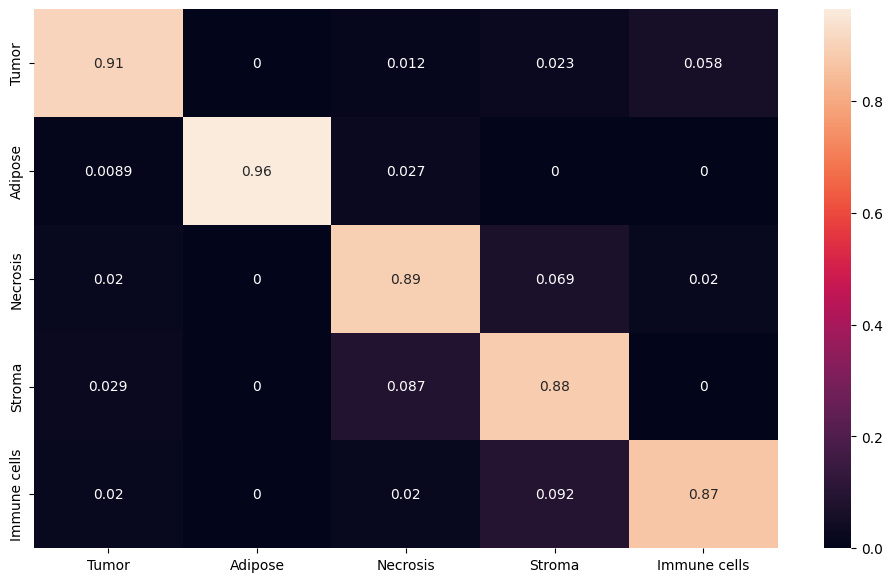}
	\mycaption{\mymarkup{Confusion matrix for region segmentation.}}
	\label{fig:result_dl_cm}
\end{figure*}

\mymarkup{In the experiment of region segmentation, we used manual annotation as shown in Figure~\ref{fig:result_regseg_trainingset}. The chosen model was resnet50. The testing target is shown in Figure~\ref{fig:result_regseg_testset}. Based on 10 fold cross-validation, we obtained the confusion matrix shown in Figure~\ref{fig:result_dl_cm}. }

\begin{figure*}[h!]
	\centering
	\begin{subfigure}[t]{0.33\textwidth}
		\centering
		\includegraphics[angle=0,width=0.9\textwidth]{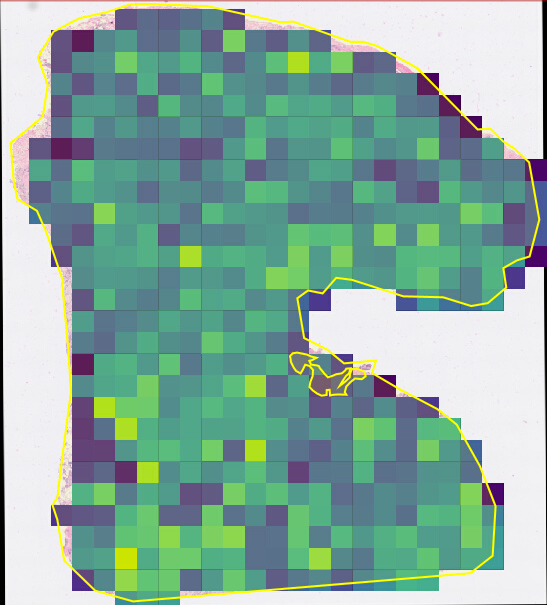}
		\caption{Tile size: 400$\mu$m}
	\end{subfigure}%
	~
	\begin{subfigure}[t]{0.33\textwidth}
		\centering
		\includegraphics[angle=0,width=0.9\textwidth]{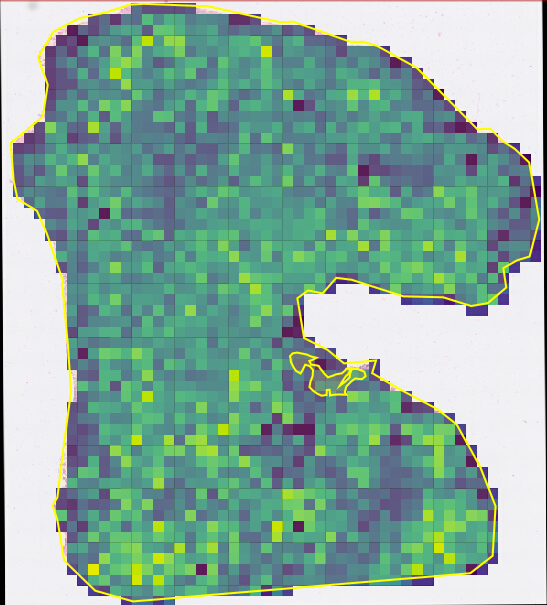}
		\caption{Tile size: 200$\mu$m}
	\end{subfigure}%
	~
	\begin{subfigure}[t]{0.33\textwidth}
		\centering
		\includegraphics[angle=0,width=0.9\textwidth]{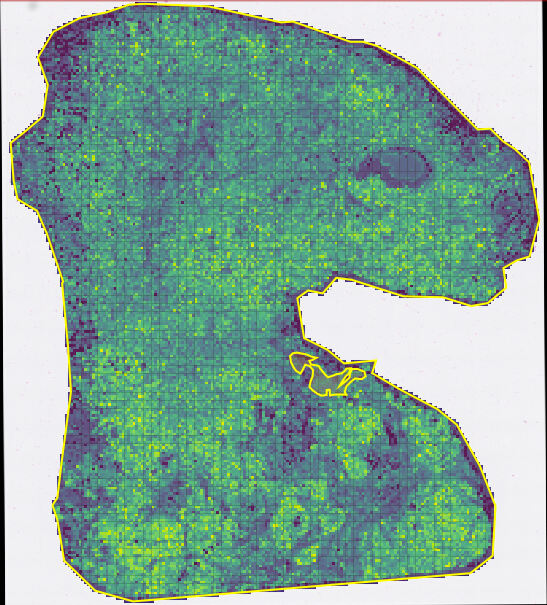}
		\caption{Tile size: 50$\mu$m}
	\end{subfigure}%
	\mycaption{\mymarkup{Image example for DL-based image region prediction. (a) the training set. (b) the testing image. (c-e) The color of heat map of each tile indicates the probability that the tile belongs to a region of tumor. Various tile sizes were presented.}}
	\label{fig:result_regseg_resolution}
\end{figure*}

\mymarkup{In addition, QuST also provides region segmentation with arbitrary tile size (aka. resolution). The higher the resolution, the longer the processing time. Figure~\ref{fig:result_regseg_resolution} shows an example of various resolutions for region segmentation.}

\section*{Data Availability}

Given the availability of H\&E images, in the experiments, we mainly used 10x~Genomics datasets. Below are the two datasets used:
\begin{itemize}
	\item FFPE Human Breast using the Entire Sample Area (\url{https://www.10xgenomics.com/datasets/ffpe-human-breast-using-the-entire-sample-area-1-standard}). The sample was 5$\mu$m section from a FFPE human breast resected tumor mass sample of Infiltrating Ductal Carcinoma, provided by Avaden Biosciences.
	\item FFPE Human Breast with Custom Add-on Panel (\url{https://www.10xgenomics.com/datasets/xenium-ffpe-human-breast-with-custom-add-on-panel-1-standard}). The sample was 5$\mu$m section from a FFPE human Infiltrating ductal carcinoma, Ductal carcinoma in situ, provided by BiolVT.
\end{itemize}

\section*{Code Availability}

The QuST is developed based on QuPath 0.5.1 and Python 3.10+ and is available under the Apache 2.0 license (\url{https://github.com/huangch/qust}). 


\section*{Author Contributions}

C.H. conceived the project, developed the algorithms and the QuPath plugin. S.L. trained and evaluated the machine learning models for image analysis, and maintained the documentation for the project. D.F. provided consultations and pathological opinions for the experiments.

C.H. wrote the manuscript with revisions from all authors. All authors approved the final version of the manuscript and agreed to submission.


Correspondence to Chao-Hui Huang.

\section*{Competing Interests }

The authors have no competing interests as defined by Nature Research, or other interests that might be perceived to influence the interpretation of the article.

\bibliography{qustwsi-v3}

\end{document}